\documentclass[
 reprint,
superscriptaddress,
showpacs,
 amsmath,amssymb,
 aps,
prb,
]{revtex4-1}

\usepackage{graphicx}
\usepackage{bm}
\usepackage{pgfplots}
\usepackage[ colorlinks,
    linkcolor={blue},
    citecolor={blue},
    urlcolor={blue}]{hyperref}
\usepackage{braket}
\usepackage[varg]{txfonts}

\begin{document}

\title{Ab-initio quantum transport through armchair graphene
 nanoribbons: streamlines in the current density}

\author{Jan Wilhelm}\affiliation{Institute of Nanotechnology, Karlsruhe Institute of Technology, D-76021 Karlsruhe, Germany}  \affiliation{Institut f\"ur Theorie der Kondensierten Materie, Karlsruhe Institute of Technology, D-76128 Karlsruhe, Germany}  

\author{Michael Walz}\email{michael.walz@kit.edu}\affiliation{Institute of Nanotechnology, Karlsruhe Institute of Technology, D-76021 Karlsruhe, Germany}  \affiliation{Institut f\"ur Theorie der Kondensierten Materie, Karlsruhe Institute of Technology, D-76128 Karlsruhe, Germany}  
\affiliation{DFG Center for Functional Nanostructures, Karlsruhe Institute of Technology, D-76131 Karlsruhe, Germany}  

\author{Ferdinand Evers} \affiliation{Institute of Nanotechnology, Karlsruhe Institute of Technology, D-76021 Karlsruhe, Germany}  \affiliation{Institut f\"ur Theorie der Kondensierten Materie, Karlsruhe Institute of Technology, D-76128 Karlsruhe, Germany}  
\affiliation{DFG Center for Functional Nanostructures, Karlsruhe Institute of Technology, D-76131 Karlsruhe, Germany}  

\vskip 0.25cm

\date{\today}

\begin{abstract}
We calculate the local current density in pristine armchair graphene nanoribbons 
(AGNRs) with varying width, $N_\text{C}$, employing a DFT-based ab-initio 
transport formalism. We observe very pronounced current patterns
(``streamlines'') with threefold periodicity in $N_\text{C}$.  
They arise as a consequence of quantum confinement in transverse flow direction. 
Neighboring streamlines are separated by stripes of almost vanishing flow.  
As a consequence, the response of the current to functionalizing adsorbates is 
very sensitive to their placement: adsorbates located within the current filaments 
lead to strong backscattering while adsorbates placed in other regions have 
almost no impact at all. 
\end{abstract}

\verb



\maketitle

\section{Introduction}

The transmission has been investigated intensely in graphene nanoribbons (GNRs) 
experimentally~\cite{TransmImpExpHan,TransmGNR,TransistGNR,DevicesGNR,
ExpGraphTrans,Cleannanoribbquantwire,Krupke1,Krupke2}, but also theoretically 
using tight-binding calculations~\cite{LiLu,CurrentPatternwithImp,
ManyimpuritiesCanada,StoneWales1,StoneWales2,AnalytImpGoeteborg,GreifswaldfromDPG} 
and first-principles approaches~\cite{RocheBNTransm,RocheChemTransm,
ImpItalien,RocheOTransm,RocheEdge,RocheBNmany,ImpFinnland,RingCurrentPaper}. 
This strong interest in GNRs is closely related to their electronic properties: 
GNRs exhibit a bandgap~\cite{energygapsgraphenenanoribb,BandgapMichelsZitat,
Michelnanoribbon} that can be tuned with the ribbon width. This makes them 
promising materials for applications, e.g., in organic opto\-electronics.  

\begin{figure}[t]
\centering
\includegraphics{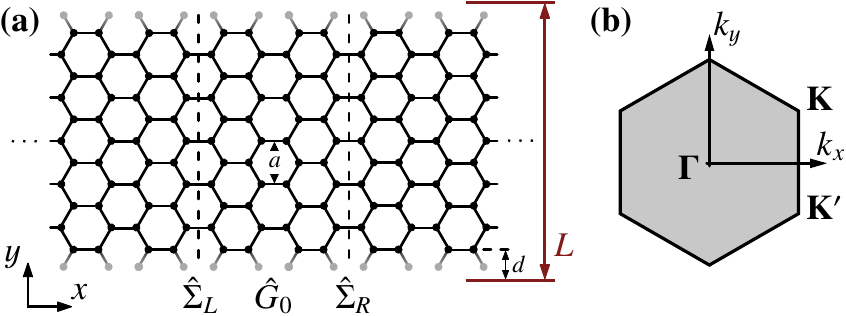}
\caption{\label{f3}(a) Structure of a hydrogen-terminated AGNR11 and (b) 
corresponding orientation of the  Brillouin 
zone of the honeycomb lattice with the $\mathbf{K}$ points 
$\mathbf{K}/\mathbf{K'}=\frac{2\pi}{3a}\left({\sqrt{3},\pm1}\right)$.}
\label{fig1}\vspace{-1em}
\end{figure}

Quite generally, the design of functional devices will
 benefit from chemical modifications of pristine ribbons, for 
instance by placing adsorbates or substituents 
like boron or nitrogen to 
achieve $p$-type or $n$-type doping. 
The electrical conductance of  functionalized 
armchair GNRs (AGNRs, see Fig.~\ref{f3})  
is typically reduced due to resonant backscattering 
with localized states caused by the 
impurity~\cite{CurrentPatternwithImp,ManyimpuritiesCanada,RocheBNTransm,
RocheChemTransm,ImpItalien,RocheOTransm,ImpFinnland,RingCurrentPaper}. 
As is well confirmed by now, the impact of a single impurity on the conductance 
is extremely sensitive to its precise placing; the conductance can drop by an order of 
magnitude when shifting the adsorbate from one carbon atom to a neighboring one. 
 Other defects, such as edge disorder~\cite{LiLu,StoneWales2} or Stone-Wales 
defects~\cite{StoneWales1,StoneWales2}, also affect the transmission 
but with a dramatically lower sensitivity to the precise defect location. 

Motivated by this peculiar situation, we simulate in this paper
the dc-current flow through pristine GNRs. 
Within tight-binding models, local (``bond'') currents are frequently 
discussed objects in the context of magnetism, e.g.~in 
Ref.~\onlinecite{WakStatesNanoribbons1999}.  Concerning experiments, 
first measurements of local dc transport properties have already been 
performed\footnote{As an example, the local electrical potential 
was measured by Kelvin probe force microscopy in graphene~\cite{ExpGoettingen,
KFPM1,KFPM2}}. In contrast, 
{systematic} theoretical  investigations of local observables are still rare even 
for tight-binding models and almost absent on the ab-initio level.~\footnote{
A publication for GNRs, that we are aware of, employs a (nearest neighbor) 
tight-binding model to investigate a ribbon with 
$N_\text{C} {=} 14$.~\cite{CurrentPatternwithImp} 
Indeed, a pronounced pattern has been reported in the bond currents but a 
conclusive explanation has not been given.  
} In this context, it is important to notice 
that patterns in bond currents are difficult to interpret
quantitatively --- in particular with respect to the intensity of the current 
modulations. It is easy to see why: the standard tight-binding model describing 
the $\pi$ electrons of conjugate carbon has only a single parameter that 
fixes the bandwidth. Without additional input, 
i.e., the explicit spe\-ci\-fi\-cation of real-space basis functions, 
a quantitative relation between bond 
currents and the physical current density 
cannot be established at all, strictly speaking.

For this reason, we set out in this work to  investigate  
the general pattern of bias-induced current flows through GNRs quantitatively 
on the ab-initio level. A systematic dependency of the flow pattern 
on the ribbon width will be presented. The results will be 
interpreted as a manifestation of (transverse) quantum confinement. 
The reported sensitivity of the conductance of AGNRs to the 
precise placement of adsorbates will be explained and also why this 
sensitivity is absent in zigzag-nanotubes. 

\begin{figure*}[t]
\includegraphics{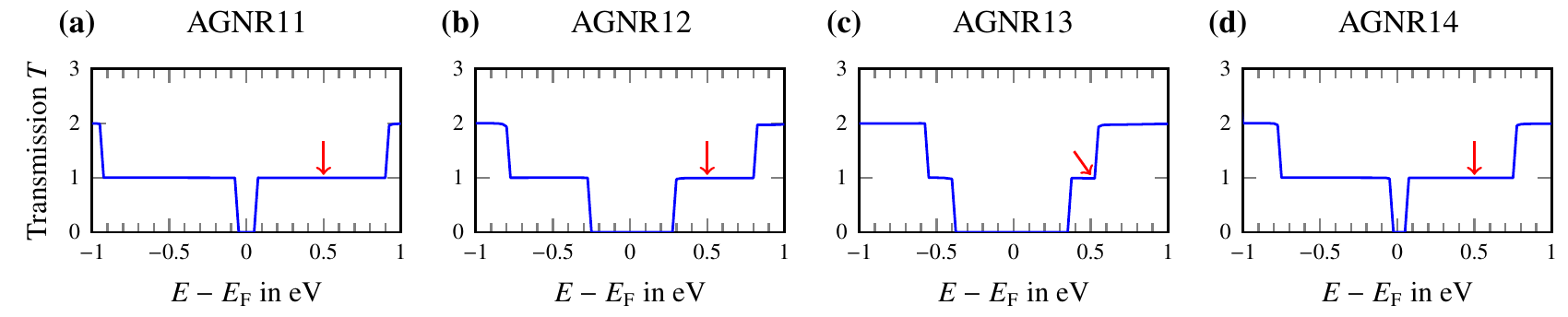}
\vspace{-0.5em}
\caption{\label{f4}
Transmission of (a) AGNR11 to (d) AGNR14. The arrows indicate the current-patterns' 
energies in Fig.~\ref{currentpatterns1114}. The reference energy, $E_\text{F}$ 
is the chemical potential of the isolated, charge-neutral species.}
\label{transmpristrib} 
\vspace{1.25em}
\includegraphics{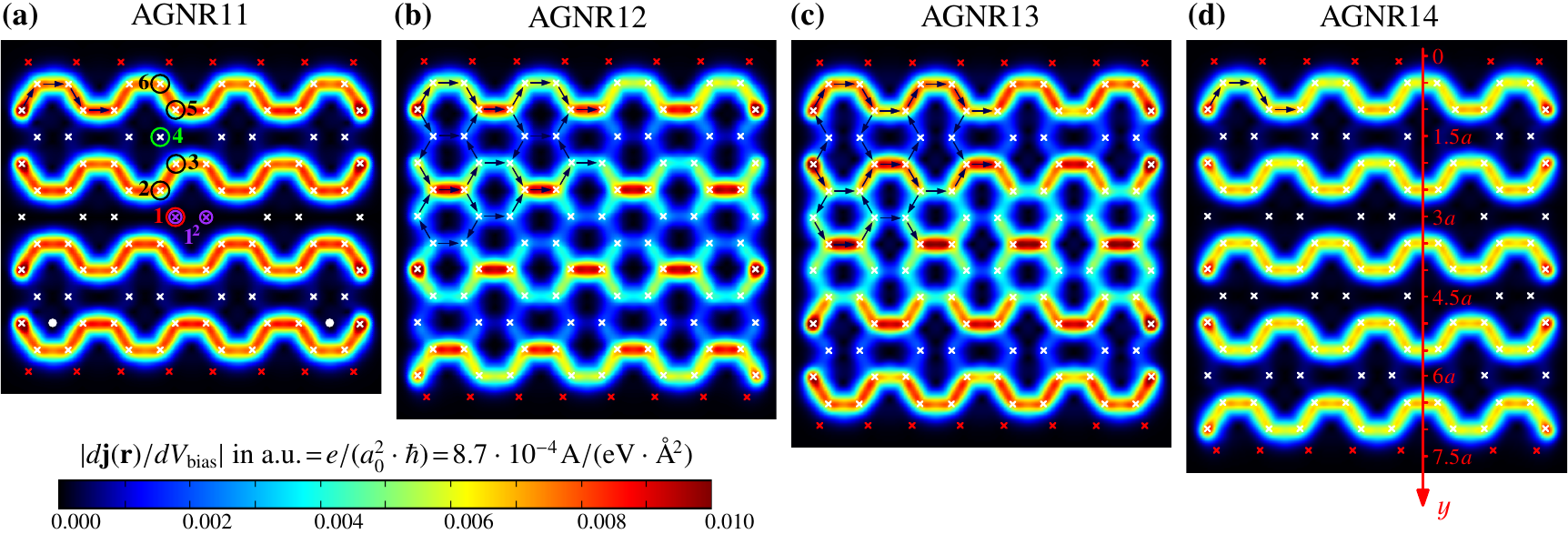}
\caption{\label{f2} Absolute value of the current density (per bias) 
associated with a fully transparent channel
in a plane $0.5\,\text{\AA}$ above the ribbon plane (exact energy 
$E=E_\text{F}+0.5\,\text{eV}$, so $T(E)=1$ for all ribbons, see 
Fig.~\ref{transmpristrib}). The current flows in horizontal direction 
as marked by the arrows. We checked that current patterns are identical 
for different energies $E$ with $T(E){=}1$. 
The calculated transmission~$T(E)$ is depicted in Fig.~\ref{transmpristrib}. 
Along the line in AGNR14, the current will be plotted  
for AGNRs$(3m{-}1)$ in Fig.~\ref{jxy}.
}\label{currentpatterns1114}
\end{figure*}

\section{Method} 
In our calculations, we are employing the AITRANSS platform, 
our DFT-based transport simulation tool~\cite{Transportcode,STM,AlexejTransport,C60TransportAITRANSS}. 
The local current density is obtained as follows~\cite{MichaelsMethod}: 
We extract the Kohn-Sham (KS) Hamiltonian out of a DFT calculation for a 
structurally non-relaxed finite-size hydrogen-terminated graphene nanoribbon 
with horizontal armchair edges
(see Fig.~\ref{f3}).~\footnote{Methodological details: TURBOMOLE 
package~\cite{Turbomole}, DFT with generalized gradient 
approximation~(GGA, BP86 functional)~\cite{GGA,BP86}
together with a contracted  Gaussian-type basis (def2-SVP)~\cite{def2SVP} 
and corresponding Coulomb-fitting basis set within the resolution of the 
identity (RI) approximation~\cite{RI}.}  Subsequently, we obtain  the (retarded) 
single particle KS-Green's function~$\hat{G}$ of a finite-size strip 
in the presence of the left and right contacts by  standard 
recursive Green's function techniques~\cite{DiVentra}: 
\begin{align}
 \hat{G}(E) = ( \hat{G}_0^{-1} -  \hat{\Sigma}_L- \hat{\Sigma}_R)^{-1}. \label{eq1}
\end{align}
The self-energies $ \hat{\Sigma}_R$ and $ \hat{\Sigma}_L$ reflect the 
presence of the reservoirs~\cite{ABC}, while $\hat G_0$ represents the bare 
KS-Green's function, see Fig.~\ref{fig1}(a). 
The local currents follow from the non-equilibrium Green's function (NEGF) formalism. 
It features the lesser Green's function with 
\mbox{$\hat{\Gamma}_\alpha=i( \hat{\Sigma}_\alpha- 
\hat{\Sigma}_\alpha^\dagger)$,}
\begin{align}
 \hat{G}^<(E) = i \hat{G}(E) \hat{\Gamma}_L(E)  \hat{G}^\dagger(E)\label{eq2}
\end{align}
that relates to the local current density (per spin): 
\begin{align}
\left.\frac{d\mathbf{j}(\mathbf{r})}{dV_\text{bias}}\;\right|_E
=\frac{1}{2\pi}\,\frac{\hbar^2}{2 m}\,\underset{\mathbf{r'}\rightarrow
\mathbf{r}}{\text{lim}}
(\boldsymbol{\nabla_\mathbf{r'}}-\boldsymbol{\nabla_\mathbf{r}})
G^<(\mathbf{r},\mathbf{r'},E)\,.
\end{align}
The factor $1/2\pi$ arises from  an inverse Fourier 
transform. 

\section{Results: Transmission and current density}
\begin{figure}[]
\centering
\includegraphics{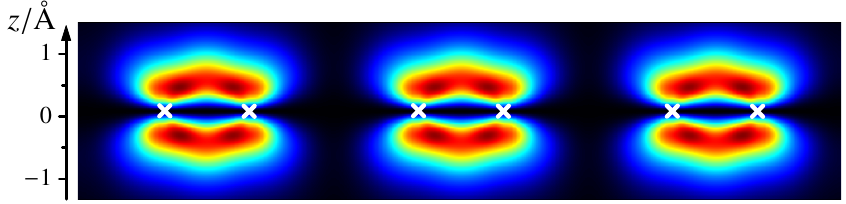}
\caption{Absolute value of the current density (per bias) 
in an AGNR11 at $T{=}1$ perpendicular 
to the ribbon plane between the  white points out of 
Fig.~\ref{currentpatterns1114}(a) using the color scale of 
Fig.~\ref{currentpatterns1114}. } 
\label{11senk}
\end{figure}

Fig.~\ref{f4} displays the transmission 
function $T(E)$ of AGNRs
of width 
$N_\text{C}$ (Nomenclature: AGNR($N_\text{C}$))~\footnote{
We calculate the transmission 
function in a Landauer-B$\ddot{\text{u}}$ttiker approach~\cite{Land,Buett}:
\begin{align*}
T(E)=\text{Tr}(\hat{G}\,\hat{\Gamma}_L\hat{ G}^\dagger 
\hat{\Gamma}_R)\,,\hspace{0.2cm}\hat{\Gamma}_\alpha
=i( \hat{\Sigma}_\alpha- \hat{\Sigma}_\alpha^\dagger)\,.
\end{align*}
Then, the total current is given by
\begin{align*}
I = \frac{2e}{h}\int_{-\infty}^\infty \text{d}E\left( f_L(E)-f_R(E) \right) T(E)
\end{align*}
with $f_\alpha(E)$ being the Fermi functions of the reservoirs.
}. 
Here, $T(E)$ simply counts the energy bands intersecting with a given energy $E$. 
The bandgap characteristic of all AGNRs with its three-fold periodicity is clearly 
seen -- mini\-mum with AGNRs$(3m{-}1)$, 
$m{\in}\mathbb{N}$  
$({\approx}0.1\,\text{eV})$.
This observation reflects a well-known behaviour~\cite{WakStatesNanoribbons1999,
NanoribbonExp,BandgapMichelsZitat,Michelnanoribbon,energygapsgraphenenanoribb}.

The corresponding current densities $d\mathbf{j}(\mathbf{r})/dV_\text{bias}$
are shown in Fig.~\ref{currentpatterns1114} and Fig.~\ref{11senk}.  
As one might expect, the current flows along chemical bonds 
following the $\pi$ orbitals. 
Due to the central node of $p_z$ orbitals, 
the current flow splits into an upper and lower sheet, see Fig.~\ref{11senk}.
Within the horizontal plane
the current density $d\mathbf{j}(\mathbf{r})/dV_\text{bias}$ 
is strongly textured, see Fig.~\ref{currentpatterns1114}. For instance, in 
AGNRs$(3m{-}1)$
the current splits into $m$  streamlines, 
with a fraction of horizontal bonds exhibiting zero flow.
For these ribbons,  the streamlines  exist in a wide energy window
whenever there is a single transparent channel, 
$T(E){=}1$.
For other ribbons, 
AGNRs$({\neq}3m{-}1)$, 
streamlines survive at the edges, but start to mix in the bulk. 
Concerning their shape, the current patterns merely reflect the symmetry of the 
underlying molecular structure. 
AGNR11 and AGNR13 exhibit a horizontal symmetry axis in the middle, while 
the others exhibit a glide reflection symmetry.

Notice, that on the level of our simulations, we see no indication 
that this distinctive threefold periodicity in current patterns 
washes out at larger values of $N_\text{C}$.
\begin{figure}[b]
\includegraphics{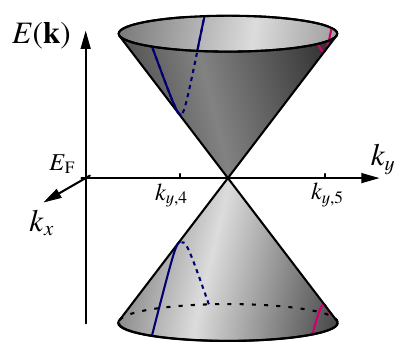}
\caption{Cone with discrete~$k_{y,n}$ points near the Dirac point~$\mathbf{K}$ 
exemplary for an AGNR12. 
 The states of this nanoribbon near the Fermi level are arranged on the  hyperbolae 
with $k_{y,4}=12/13\,K_y$ and  $k_{y,5}=15/13\,K_y$ due to Eq.~(\ref{knprec}). 
So, $k_{y,4}$ lies uniquely closest to the Dirac point, $\tilde n {=}4$. }
\label{diracpoint}
\end{figure}

\section{Discussion: Selection rules}
In order to explain the current pattern, 
we invoke a standard zone-folding argument for graphene, 
here applied to AGNRs. 
Hard-wall boundary conditions in transverse current direction, $y$, 
imply a selection rule: $k_{y,n} = {n\pi}/{L},\linebreak n\in\mathbb{N}$. 
A natural choice is 
$L=(N_\text{C} + 1)a/2$~
leading to 
\begin{align}
k_{y,n} 
=
 \frac{\pi}{(N_\text{C} + 1)a/2}
\cdot  n 
= 
\frac{2\pi}{3a}
\cdot\frac{3n}{N_\text{C}+1}\,\label{knprec},
\end{align} 
with $a{=}2.46\,\text{\AA}$ 
being the graphene lattice constant.~\footnote{There are 
two ways to motivate our choice of 
$L{\propto} N_\text{C}{+}1$ 
(rather than e.g.~$L{\propto} N_\text{C}$).
First, in order to implement hard-wall BC in a tight-binding model, 
one would add one further site at 
each transverse edge  with an infinite on-site potential, 
consequently $L{=}(N_\text{C}{+}
 1)a/2$.
Second, a more realistic model would be a box with 
length~$L{=}(N_\text{C}{-}1)
a/2$ 
and finite-potential walls with the work function~$\Phi$ of the nanoribbon 
as potential height. In this model, the $\pi$ electrons are located between
 the two outermost box atoms and the wavefunctions decay outside the box 
exponentially with the tunneling length scale~$\lambda_\text{T}$. This 
length scale should approximately coincide with the distance~$d$ between 
the outer carbon atom and the border of the box (see Fig.~\ref{fig1})      
and is calculated using~$\Phi{=} 4{.}0\,\text{eV}$ 
of DFT calculations for ribbons and compared 
to~$a{=}\text{2.46}\,\text{\AA}$:
$
\lambda_\text{T}={\hbar}/{\sqrt{2m\Phi}}=0{.}98\,\text{\AA}\approx {a}/{2}\,.
$
In the model with infinite-potential walls, $d$ must be 
chosen in the order of 
$\lambda_\text{T}$ and $d{=}
a/2$ 
fulfills this condition.
}

For every ribbon with width $N_\text{C}$, 
there is a unique (see Fig.~\ref{diracpoint}) integer~$\tilde{n}$
that minimizes the distance $|k_{y,n}-K_y|$, 
$K_y = \frac{2\pi}{3a}$.
The implications of similar selection rules for the spectrum 
were already investigated and discussed by previous 
authors~\cite{WakStatesNanoribbons1999,BreyFertig,RevElStatGr}. 
Here, our interest is in the consequences for spatial properties, 
in particular the wavefunction's nodal structure 
and the node placing with respect to the carbon lattice. 
\begin{figure}[]
\centering
\includegraphics{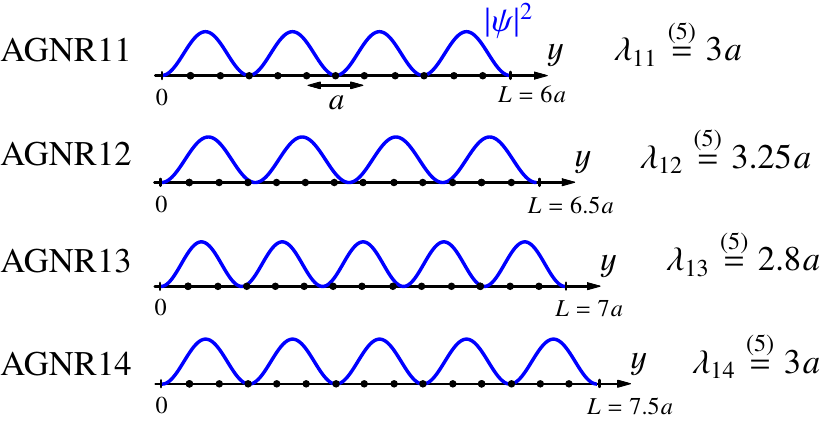}
\caption{\label{f6} Box with transversal 
density~$|\psi|^2{\propto}\sin^2(k_{y,\tilde n}y)$ 
of electronic states in a box with a wavelength according to Eq.~(\ref{11}). 
Charge carriers near the Fermi level have to  jump to states with such an 
envelope  when crossing an AGNR. }\label{wavef}
\end{figure}
For this reason, we focus on the wavelength
$\lambda_{N_\text{C}}=2\pi/k_{y,\tilde{n}}$:
\begin{align}
\lambda_{N_\text{C}} = 3a \cdot
\left\{
\begin{array}{ll}
1  &\text{for}\;  N_\text{C} = 3 m-1 \\[0.3em]
1 + 1/N_\text{C} &\text{for}\;  N_\text{C} = 3 m \\[0.3em]
1 - 1/(N_\text{C}+2) &\text{for}\; N_\text{C} = 3 m+1 
\end{array}
\right..\label{11}
\end{align}

\begin{figure}[b]
\includegraphics{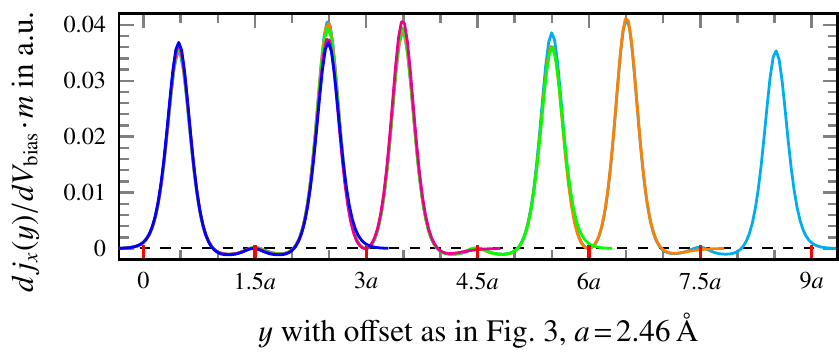}
\caption{$x$-component of the current density (per bias) scaled by the number of 
streamlines~$m$ at $E{=}
E_\text{F}{+}0.5\,\text{eV}$  
along the line sketched in the AGNR${\color{orange}{14}}$ in 
Fig.\,\ref{currentpatterns1114}(d) and analogously 
for  AGNRs$(3m$\text{$-$}1) with 
$N_\text{C}{=}{\color{blue}{5}},
{\color{magenta}{8}},{\color{green}{11}}$ and $\color{cyan}{17}$. 
 Red ticks: node position, see Fig.~\ref{currentpatterns1114} and~\ref{wavef}. 
Negative currents near nodal lines indicate small current eddies with backflow. 
}
\label{jxy}
\end{figure}

As can be seen from Fig.~\ref{wavef},
the boundary conditions imply a partial mismatch of the wavefunction 
extrema with the carbon lattice. 
The exceptions are
AGNRs$(3m$\text{$-$}1), 
where the nodes of the wavefunction coincide with the carbon sites
as a consequence of the perfect fit  
of $\lambda_0{=}3 a$ into the box of 
AGNRs$(3m$\text{$-$}1). 
Hence, charge carriers have no probability amplitude on these carbon sites and
 therefore the connecting chemical bonds cannot carry current. Since there 
 are $m{-}1$ disconnecting bonds, 
the number of streamlines is~$m$. 
We recover the salient features of Fig.~\ref{f2}. 
Moreover, the simple particle-in-the-box picture predicts 
(i) that the shape of the current envelope is the same for each streamline
and (ii) that the shape is independent of the ribbon width~$N_\text{C}$. 
Both predictions are seen to be confirmed in Fig. \ref{jxy}.  

The wavefunction's nodes of non-matching ribbons, 
$N_\text{C}{\neq}3m{-}1$,
are seen to be displaced from the carbon sites in Fig.~\ref{f6}, 
with small displacements near the edges and 
shifts of order of the lattice constant $a$ in the bulk. Correspondingly, the 
current density shows streamlines near the edges but a less pronounced 
valley structure in the bulk, consistent with earlier observations in 
Fig.~\ref{f2}.

\begin{figure}[t]
\centering
\includegraphics{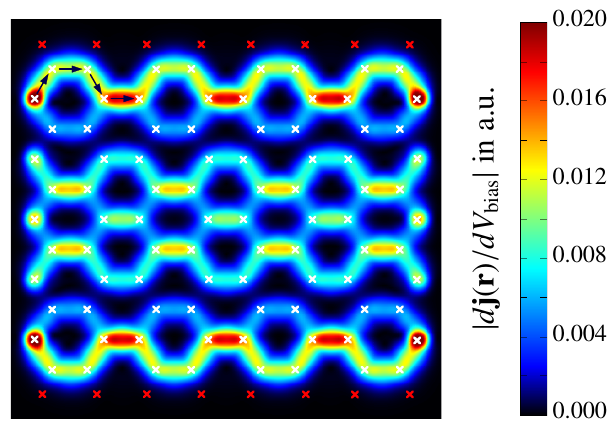}
\caption{ Absolute value of the current density 
in an AGNR11 with two transmission channels, $T{=}2$ (again, in a plane 
$0.5\,\text{\AA}$ above the ribbon, exact energy 
$E{=}E_\text{F}{+}1.0\,\text{eV}$). The observed pattern follows from
a superposition of the channel at $T{=}1$ in Fig.~\ref{currentpatterns1114}
and an additional channel with a transversal wavelength~$\lambda{\neq}3a$.
Similarly to AGNRs$({\neq}3m{-}1)$ at $T{=}1$, streamlines survive at 
the edges, but start to mix in the bulk.
\label{fig7}}
\end{figure}

Further remarks:
(a) All electronic states in a given band (e.g.~see Fig.~\ref{diracpoint}) 
exhibit the same $k_y$-component, i.\,e. share the same nodal structure. 
Therefore, they inherit the same current pattern which explains the robustness of 
the observed patterns, e.g., against shifting the Fermi energy. 
(b) 
If more than a single band contributes to the current, the total current will 
be a superposition of all  bands with streamlines that in general don't 
necessarily match. In this case, we observe a more complicated pattern, see 
the example in Fig.~\ref{fig7}. 
(c) The energy interval~$\Delta E$ with single-channel transport 
($T(E){=}1$) decreases with the ribbon 
width:
\begin{align}
\Delta E=\frac{A}{N_\text{C}}
\hspace{0.5cm}\text{with}\hspace{0.5cm}
A=25\,\text{eV}\,,
\end{align}
see Appendix~\ref{sec:AppB} for more details.
(d) The nodal structure of the wavefunctions is also displayed in the local density
of states. Therefore, the nodal pattern may also be observed in 
STM experiments as simulations indicate~\cite{STMRibbon}, see Appendix~\ref{sec:AppA}.

\section{Application: adsorbate placing and transport}
\begin{figure}[t]
\centering
\includegraphics{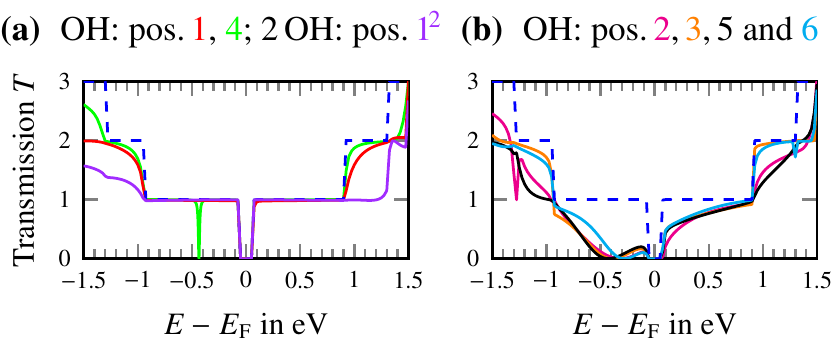}
\vspace{-0.8em}
\caption{\label{f7} Transmission of defected AGNR11 and the 
pristine case (dashed) for different OH positions out of  
Fig.~\ref{currentpatterns1114}. Left (right): OH is positioned 
outside (inside) a streamline.}
\label{transmdefrib}\vspace{-0.75em}
\end{figure}

In the final section, we apply our results in order to explain 
earlier findings on the transport properties of functionalized AGNRs. 
It is well known~\cite{RocheBNTransm,RocheChemTransm,ImpItalien,
RocheOTransm,ImpFinnland} 
that the impact of an impurity 
(adsorbate that promotes a carbon atom from $sp^2$ to $sp^3$ hybridization) 
on the transmission is very strongly site dependent. 
We confirm this observation in Fig.~\ref{f7}. 
It shows that the transmission function of an AGNR11
(structurally relaxed) in the presence of a single 
OH group is extremely sensitive to placement of the adsorbate. 
Shifting by one lattice site can change the transmission 
by orders of magnitude in a wide energy window.

When analyzing this observation in terms of streamlines, it has a very 
simple intuitive explanation. If streamlines don't touch the adsorption 
site (at pos.~1 and~4 in Fig.~\ref{currentpatterns1114}), the transmission 
is hardly affected by the OH group.~\footnote{Arguments based on 
wavefunction symmetries and vani\-shing wavefunction overlap matrix 
elements have been given that with the adsorption site~1 backscattering 
is completely suppressed~\cite{LiLu,RocheBNTransm}}
Our observation implies that even a finite concentration of adsorbates 
leaves the transmission  invariant as long as they are placed in 
regions of zero flow (see Fig.~\ref{transmdefrib} for the case of 2\,OH). 
 This allows an impurity concentration up to $1/3$ without significant 
 influence on  the transmission. By contrast, when placing an impurity 
 right into a streamline (pos.~2, 3, 5 and~6 in 
 Fig.~\ref{currentpatterns1114}), the transmission is strongly 
 perturbed, see Fig.~\ref{transmdefrib}.

Since the current pattern is a quantum confinement effect, one might wonder 
about the fate of the placing sensitivity after changing 
from hard-wall to periodic boundary conditions, i.e. from 
AGNRs to (zigzag) carbon nanotubes. Clearly, 
due to the transverse periodicity all carbon atoms are equivalent and therefore  
the  tube's transmission cannot depend on the impurity placing, 
consistent with findings in the literature~\cite{TransmTubes,RocheBNTransm}.

\section{Conclusions}
In conclusion, we study the local current density per bias voltage, 
$d\mathbf{j}(\mathbf{r})/dV_\text{bias}$,  
in pristine armchair graphene nanoribbons 
with transport density functional theory. Our most important result is that 
$d\mathbf{j}(\mathbf{r})/dV_\text{bias}$ shows pronounced streamlines; 
the pattern exhibits a threefold periodicity  in the width of the ribbon.

We explain the effect as a consequence of quantum confinement in 
transverse current direction. Due to streamlines, there is a strong 
sensitivity of the current response to the local placement of adsorbates. 
This sensitivity was well known before,  and our results can  provide 
an intuitive understanding of it. Finally, we mention that our results 
can also be understood as a manifestation of strong spatial structure 
in the scattering states of mesoscopic devices. 
We expect, that the structural elements --- ``current filaments'' --- that
were observed in this study are a generic feature of transport 
through meso- and nano-devices that has hardly been touched upon by now.

\begin{acknowledgments}
We thank C. Seiler and A. Bagrets for stimulating discussions and express
our gratitude to the Simulation Lab NanoMicro, especially to I. Kondov, for
computational support. The authors gratefully acknowledge the computing time
granted by the John von Neumann Institute for Computing (NIC) and provided
on the supercomputer JUROPA at J\"ulich Supercomputing Centre (JSC). 
\end{acknowledgments}

\appendix
\section*{Appendix: Energy range of streamlines and STM images }
In Appendix~\ref{sec:AppB}, the energy interval where streamlines appear is discussed 
briefly. Appendix~\ref{sec:AppA} provides simulation results for the energy-resolved 
equilibrium local  electron density of AGNRs which can in principle be detected by a 
scanning tunneling microscope.

\section{Energy window for observation of streamlines}\label{sec:AppB}
As already mentioned in the body of the paper, the current pattern and 
the density pattern arrange in streamlines in AGNRs(3$m{-}1$) only at 
energies~$E$ with a single, fully transparent channel. 
For energies  farther away from the Fermi level, with two or more 
current channels, the patterns are more complicated. 
The energy range~$\Delta E$ with $T(E)\hspace{0.1em}{=}\hspace{0.1em}1$ 
 is equal to the distance between 
the second upper and the second lower band minus the bandgap, see 
Fig.~\ref{transmpristrib} and Fig.~\ref{diracpoint}.
One expects a $1/N_\text{C}$ behaviour for this energy interval~$\Delta E$, 
since the discrete $k_{y,n}$ scale with~$1/N_\text{C}$. 
This behaviour is checked in Table~I for AGNRs(3$m{-}1$).

We see a $1/N_\text{C}$ behaviour: The product of $\Delta E$ and 
$N_\text{C}$ is approximately constant for bigger ribbons, 
while it is deviating strongly for smaller ribbons possibly caused by 
additional finite-size effects; 
Summarizing: 
\begin{align}
\Delta E = A\cdot \frac{1}{N_\text{C}}\hspace{0.5cm}\text{with}
\hspace{0.5cm}A\approx 25\,\text{eV}
\end{align}
for AGNRs(3$m{-}1$) with $m\hspace{0.1em}{\ge}\hspace{0.1em}4$. 
In this energy interval, the streamlines with their characteristic nodes 
are expected to be detectable by STM. In particular, the energy window  
vanishes for the bulk 
limit~$N_\text{C}\hspace{0.1em}{\rightarrow}\hspace{0.1em}\infty$.

\begin{table}[]
\begin{ruledtabular}
\begin{tabular}{lll}
 ribbon width~$N_\text{C}$ & Energy range 
$\Delta E$ & $\Delta E\cdot N_\text{C}$ in eV \\[0.2em]
& with $T\hspace{0.1em}{=}\hspace{0.1em}1$ in eV
\\[0.5em]
\colrule
5 &2.63$^\text{a}$ & 13.2 \\[0.2em]
8 &2.20$^\text{a}$ & 17.6 \\[0.2em]
11 &1.75$^\text{a}$ & 19.3 \\[0.2em]
14 &1.48$^\text{a}$ & 20.7 \\[0.2em]
17 &1.20$^\text{a}$ & 20.4 \\[0.2em]
20 &1.30\cite{RocheBNTransm} & 26.0 \\[0.2em]
35 & 0.75\cite{RocheBNTransm} & 26.3 \\[0.2em]
41 & 0.60$^\text{b}$ & 24.6 \\[0.2em]
44 & 0.60\cite{ImpItalien}  & 26.4
\\[0.5em]
 \end{tabular}
\caption{Testing the dependency of $\Delta E$ and $N_\text{C}$ for AGNRs(3$m{-}1$). 
$\Delta E$ is defined as the energy range when only one band is present. 
All calculated values of~$\Delta E$ are obtained by ab-initio calculations.
In this energy interval~$\Delta E$, the streamlines appear. If the 
dependency $\Delta E(N_\text{C})$ obeys an inverse correlation 
$\Delta E\hspace{0.1em}{=}\hspace{0.1em} A/N_\text{C}$, 
the product 
$\Delta E\hspace{0.15em}{\cdot}\hspace{0.05em} N_\text{C}$ 
will be~$A$ for all~$N_\text{C}$. 
One shouldn't attach importance to the exact numerical 
values since they are strongly functional/method dependent.
}
\footnotetext{The ranges were extracted from transmission curves as in Fig.~2.}
\footnotetext{We applied the FHI-aims packages~\cite{FHIaims} using PBE 
functional for the DFT calculation employing tier1 basis set.}
\vspace{-0.8em}
\end{ruledtabular}
\end{table}

\begin{figure*}[]
\centering
\includegraphics{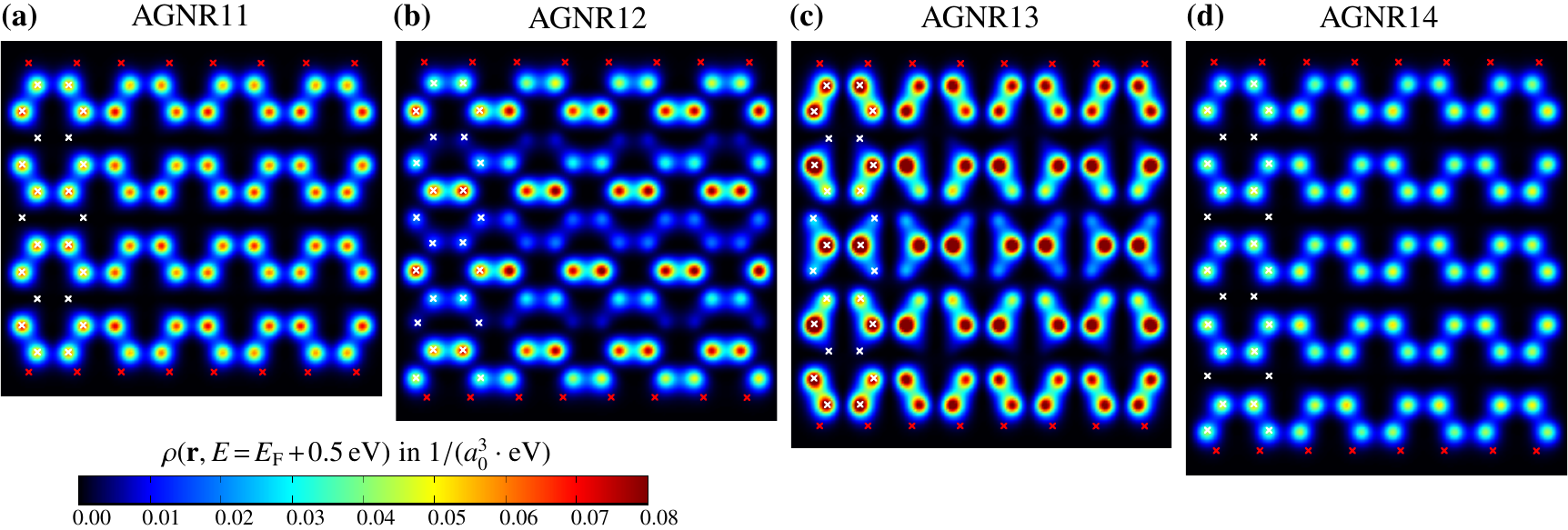}
\caption{LDoS in a plane $0.5\,\text{\AA}$ 
above the ribbon plane 
(exact energy $E=E_\text{F}+0.5\,\text{eV}$, so $T(E)=1$ for all ribbons, 
see Fig.~2). 
We checked that the LDoS patterns are identical for energies~$E$ 
with $T(E)\hspace{0.1em}{=}\hspace{0.1em}1$ except for their amplitude 
being dependent on the exact energy. Due to the close AGNR13's van 
Hove singularity (at 
${\approx}\hspace{0.1em}E_\text{F}\hspace{0.1em}{+}\hspace{0.1em}
\text{0.4\hspace{0.1em}eV}$), the LDoS at $E\hspace{0.1em}{=}
\hspace{0.1em}E_\text{F}\hspace{0.1em}{+}\hspace{0.1em}
0.5\hspace{0.1em}\text{eV}$ 
is enhanced in AGNR13 compared to the other AGNRs.  \label{neqdens} }
\vspace{-1em}
\end{figure*}

\section{STM images}\label{sec:AppA}

\subsection{Results: Local equilibrium density of states}

The energy-resolved local equilibrium density of states (LDoS) in the
presence of the leads
is calculated as~\cite{DiVentra,MichaelsMethod}
\begin{align}
\rho(\mathbf{r},E) =-\frac{1}{\pi}\,\text{Im}\,G(\mathbf{r},\mathbf{r},E)\,,
\end{align}
where $G(\mathbf{r},\mathbf{r},E){=}\braket{\mathbf{r}|
 \hat{G}(E)|\mathbf{r}}$,
see Eq.\,\eqref{eq1}.
The simulation results are shown in Fig.~\ref{neqdens}: 
The LDoS at 0.5\,eV above the Fermi level shows strong texturing 
not only in the direction transverse to the current flow 
but also in longitudinal direction parallel to the streamlines
(compare to Fig.~\ref{currentpatterns1114}).  
As one would expect, also the LDoS inherits the nodal structure of transverse 
wavefunctions near the Fermi level. For instance AGNR11 and AGNR14: 
the density on every third carbon atom (in transverse direction) vanishes
consistent with  Fig.~\ref{f6}.

Notice that in contrast to the density the current obeys the continuity 
equation,~$\boldsymbol{\nabla}\cdot\mathbf{j}=0$. 
Hence, texturing in longitudinal direction is suppressed for the current 
resulting in streamlines that cannot be observed in the LDoS.

\begin{figure}[]
\vspace{3.5em}
	\centering
\includegraphics{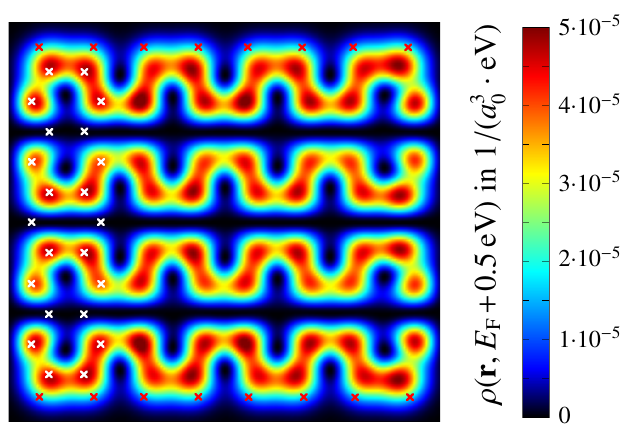}
\caption{LDoS of an AGNR11 in a plane $2\,\text{\AA}$ above the ribbon plane 
at $E\hspace{0.1em}{=}\hspace{0.1em}E_\text{F}\hspace{0.1em}{+}
\hspace{0.1em}0.5\hspace{0.1em}\text{eV}$. }\label{neqdens11gz}
\end{figure}

\subsection{Pseudostreamlines in STM images at zero (in-plane) current flow}

In the simplest picture (Tersoff-Hamann theory~\cite{TersoffHamann}), 
a scanning tunneling microscope (STM) detects the energy-resolved local
electron density of states (LDoS). 
Hence, the STM would detect patterns similar to Fig.~\ref{neqdens} 
if it operated in a constant-height mode 
with a tip-distance~$z$ very close to the substrate: 
$z\hspace{0.1em}{=}\hspace{0.1em}0.5\hspace{0.1em}\text{\AA}$. 
With increasing the ribbon-tip distance, the STM resolves less and less 
features of the $\pi$-system so that the longitudinal texturing 
washes out, see Fig.~\ref{neqdens11gz}. 
In contrast, the transverse nodal structure survives since even at large  
distances the sign change of the wavefunction can be detected. 
Therefore, one could expect that STM-images taken at larger distances show 
an LDoS patterned in a streamline-type manner (``Pseudostreamlines'').

The simulated STM image out of Fig.~\ref{neqdens11gz} shows this feature. 
It is in perfect agreement with results of earlier 
authors~\cite{STMRibbon}, who have explained this pattern in a much more 
complicated way, though. They apply Clar's theory, a rule originating from  
organic chemistry~\cite{Clar0,Clar1} that 
relies on a proper placing of (many) double bonds.

\subsection{Experiments}

We are not aware of experimental STM images of AGNRs(3$m{-}1$) that would be 
showing streamlines or pseudostreamlines probably, 
because the control  over the edge geometry is still an experimental challenge. 
Edges irregularities lead to  electronic states with node structures strongly 
deviating  from those of clean AGNRs(3$m{-}1$). 

There are indications of the existing of standing wave patterns in other carbon 
based conjugate matter, namely in the fullerenes. Since also those may be 
thought of as graphene derivates, one would expect a similar wavelength 
~$\lambda\hspace{0.1em}{=}\hspace{0.1em}3\hspace{0.05em}a
\hspace{0.1em}{=}\hspace{0.1em}7.4\,\text{\AA}$ 
to appear there as well. 

Indeed, consider recent  
STM experiments on the fullerenes~C$_{58}$ and C$_{60}$.~\cite{Nakaya,C58STMexp} 
For a single fullerene, (low bias) STM images  
show an structureless circular spot, roughly consistent with the 
absence of the analogue of hard-wall boundary conditions --- roughly
similar to (zigzag) carbon nanotubes discussed above. 
Boundary conditions removing the rotational symmetry are realized with a formation 
of a chemical bond between two different fullerene cages. Then, STM images show 
a stripe pattern with a characteristic 
wavelength~${\approx}\hspace{0.1em}\lambda$. 
Given the discussion above, it is not surprising, perhaps, that 
this wavelength transfers from AGNR to fullerene electronic states near 
the Fermi level since fullerenes and nanoribbons are both a certain kind 
of derivate of graphene.

\bibliography{Literature}

\end{document}